\documentstyle[12pt,aaspp4]{article}
\def\clock{\count0=\time \divide\count0 by 60
     \count1=\count0 \multiply\count1 by -60 \advance\count1 by \time
     \number\count0:\ifnum\count1<10{0\number\count1}\else\number\count1\fi}

\begin{document}
\title{$\Omega_0$ from the Apparent Evolution of Gas Fractions in Rich
Clusters of Galaxies}
\author{Rebecca Danos}
\affil{Whitin Observatory, Wellesley College, Wellesley MA 02181, USA}
\author{and Ue-Li Pen}
\affil{Harvard-Smithsonian Center
for Astrophysics, 60 Garden St., Cambridge, MA 02138, USA}

\newcommand{\etal}{{\it et al. }}
\newcommand{\beq}{\begin{equation}}
\newcommand{\eeq}{\end{equation}}

\begin{abstract}
We apply a unique gas fraction estimator to published X-ray cluster
properties and compare the derived gas fractions of observed
clusters to simulated ones.  The observations are consistent with a
universal gas fraction of $0.15\pm 0.01 h_{50}^{-3/2}$ for the low
redshift clusters that meet our selection criteria.  The fair sampling
hypothesis states that all clusters should have a universal constant
gas fraction for all times.  Consequently, any apparent evolution
would most likely be explained by an incorrect assumption for the
angular-diameter distance relation.  We show that the high redshift
cluster data is consistent with this hypothesis for $\Omega_0<0.63 $
(95\% formal confidence, flat $\Lambda$ model) or $\Omega_0<0.60$ (95\%
formal confidence, hyperbolic open model).  The maximum likelihood
occurs at $\Omega_0=0.2$ for a spatially flat cosmological constant
model.
\end{abstract}

\section{Introduction}

It has been proposed that clusters of galaxies should be a fair sample
of baryonic matter (White \etal 1993).  Rich clusters form through the
gravitational collapse of the matter within a 15-30 Mpc diameter
volume, where the force of gravity acts equally on all
non-relativistic forms of matter.  The richest clusters have
temperatures above 10 keV, which corresponds to velocity dispersions
in excess of 1000 km/sec.  Most non-gravitational processes do not
appear to affect the bulk of matter at comparable energies, and so we
would expect the gas and dark matter to collapse into objects where
they are fairly represented.  This is confirmed in simulations using a
large variety of techniques (Frenk
\etal 1998), where it is found that within the virial radius the gas
and dark matter are indeed equally represented with deviations of only
10\%.  Since clusters of galaxies are observed at cosmological
distances and are spatially resolved objects, this opens the
possibility of directly measuring angular diameter distances if the
gas to dark matter ratio were known in advance (Pen 1997).  In this
paper we compare observed and simulated cluster properties and we
estimate the errors in our methods.

\section{Method}

Observable parameters can be converted to physical quantities by using
many simplifying assumptions.  Our models most strongly depend on the
assumptions of hydrostatic equilibrium and spherical symmetry.  These
assumptions give sufficient information to solve the equations of
hydrostatic equilibrium and thus to measure the ratio of the gas mass
to the total mass within the virial radius, a value also corresponding
to its global ratio (White \etal 1993).  Currently, most astrophysical
observations either measure spatially resolved X-ray surface
brightness or broad aperture X-ray spectra, but usually not
simultaneously.  As a result, we can not directly measure total
masses.  As a first approximation, we shall therefore assume that the
gas traces the total mass at all radii, which reduces the free
parameters to allow a unique solution of the gas fraction in terms of
the three observable parameters: temperature, angular size and X-ray
flux (Pen 1997).  These quantities are usually translated into
physical size and luminosity by assuming some luminosity and
angular-diameter distance relation $d_L(z),\ d_A(z)$.  We can estimate
the error induced by these assumptions by applying this scheme to the
simulated clusters.

The Jones and Forman (1998, hereafter JF98, see also David \etal 1993)
catalog published the X-ray properties parameterized by the isothermal
$\beta$ model where the gas distribution is modeled as $\rho_g =
\rho_0 (1+r^2/r_c^2)^{-3\beta/2}$.  In our model, the observables will
be the total X-ray luminosity $L_X$ derived from the observed flux $F$
\beq
L_X=4\pi d_L^2F= 4\pi \epsilon_{\rm ff} T^{1/2} n_0^2 r_c^3 \int_0^\infty
(1+u^2)^{-3\beta} u^2 du
\eeq
in cgs units.  $\epsilon_{\rm ff}\sim 1.94\times 10^{-27}$ is the
effective free-free emissivity coefficient, which includes an average
Gaunt factor of $1.25$ and a hydrogen-helium mixture with 24\% Helium
by mass (Spitzer 1978).  The central electron density, $n_0$ from here
on in units $10^{-3}$cm$^{-3}$, can be solved by
\beq
n_0^2 = {L_{44} \Gamma[3\beta]\over 11.4 r_c^3 \sqrt{T}
\Gamma[3\beta-3/2]}
\eeq
in terms of $L_{44}$, the X-ray luminosity in units of $10^{44}$
erg/sec, the core radius $r_c$ in Mpc, $T$ in keV, and $\Gamma$
functions.  The gas fraction $f_g$ is then given by
\beq
f_g={9.37 H(\beta) n_0 r_c^2\over T}
\label{eqn:fg}
\eeq
where $H(\beta) \sim 0.057 (\beta-4/7)^{-0.787}$ (Pen 1997).  If the
absolute distance is unknown up to a linear parameter, for example
$h_{50}$, the gas fraction in (\ref{eqn:fg}) scales as
$h_{50}^{-3/2}$.

\section{Simulations}

We apply equation (\ref{eqn:fg}) to simulated clusters, in which we
know the global gas-to-mass ratio, to estimate the systematic and
statistical errors induced by the assumptions of spherical symmetry,
hydrostatic equilibrium, and gas-traces-mass.  We ran several
simulations with different volumes, resolutions, and cosmological
parameters, and we applied the observational procedure to each set of
simulations.  We list the simulation parameters in table
\ref{tbl:sim}.  Several of the simulations were previously published
(Pen 1998b), and a new higher resolution $256^3$ CDM run was added for
this analysis.  All simulations are performed with the
Moving-Mesh-Hydro code (Pen 1998a) on an Origin 2000 at the National
Center for Supercomputing applications.  The following physical
assumptions are imposed: 1. the universe is assumed to consist of two
dynamical components, dark matter and baryons in the form of an
ideal gas;  2. the dark matter is initially cold and follows the
collisionless Boltzmann equation, while the gas is described by the
Euler equations in an expanding universe.  Only simulation PREHEAT
incorporates non-gravitational processes, which simulate the worst case
scenario of energy injection.  1 keV of energy per nucleon is
instantaneously injected at a redshift of $z=1$, and the subsequent
evolution again follows Euler's equations.  This should model the net
effect of heat injection from supernova heating, which would also be
one of the consequences of cooling.

\begin{table}
\begin{center}
\begin{tabular}{|l|l|l|l|l|l|l|}
\hline
model& $\Omega_0$&$\Omega_\Lambda$&$\sigma_8$ &$h$&L($h^{-1}$ Mpc) 
&$\Delta x (h^{-1}$kpc)\\
\hline\hline
CDM  &   1   &   0       &       0.6     & 0.5   &128 & 50\\
PREHEAT& 1   &   0       &       0.5     & 0.5   &80  & 62\\
OCDM &   0.37&   0       &       1       & 0.7   &120 & 94\\
LCDM &   0.37&   0.63    &       1       & 0.7   &120 & 94\\
\hline
\end{tabular}
\end{center}
\caption{Simulations used for this study.  All simulations except CDM
used a $128^3$ grid with $256^3$ particles, while the previous used
$256^3$ grid points.  All models have a baryonic fraction $\Omega_b h^2
=0.0125$ and compression limiter $c_{\rm max}=10$.  The effective
grid spacing in virialized objects is $\Delta x$ (Pen 1998a).  We used a
periodic box of length $L$.}
\label{tbl:sim}
\end{table}

\begin{figure}
\plotone{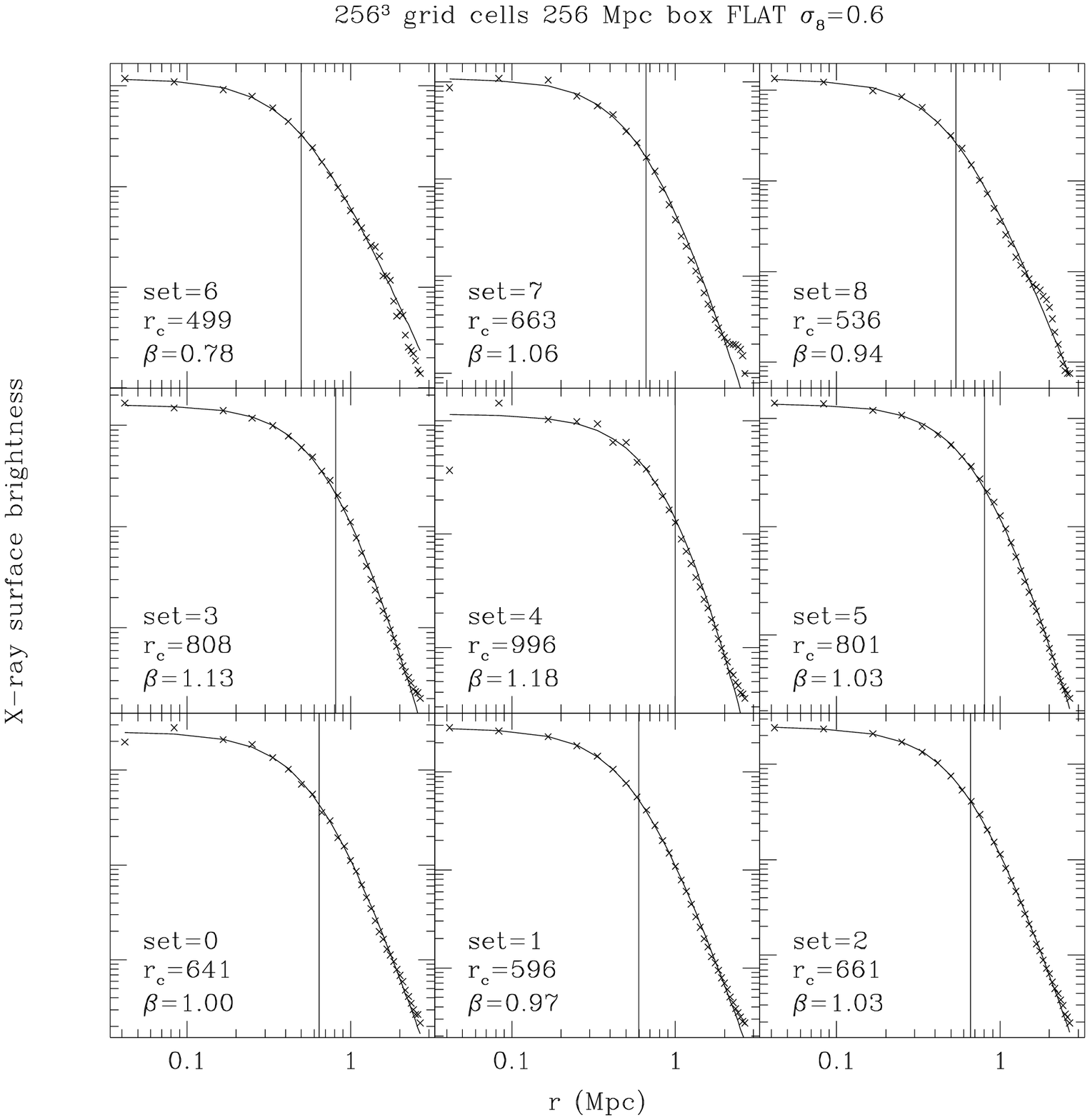}
\caption{The radial profiles fitted for simulated $\Omega=1,\ h=0.5$
clusters.  The $\beta$-model provides a very good fit for the
simulated projected X-ray surface brightness.  $r_c$ is in kpc, and
is marked by the vertical line in each panel. Each
row contains the three projections of a single cluster.}
\label{fig:q256}
\end{figure}

We first approximate the soft X-ray observations by computing the
projected density-square weighted surface brightness $\Sigma=\int
\rho^2 dz$.  In the soft X-rays, the temperature dependence is very
weak, making this a good approximation.  We then find the local maxima
in the projected 2-D X-ray image.  The surface brightness is binned
radially centered on the local maxima, and we fit a 1-D surface
brightness profiles with two free non-linear parameters $r_c$, and
$\beta$ (Press \etal 1992).  The results of the radial binning for the
highest resolution model are shown in figure \ref{fig:q256}, where the
$\beta$-model is seen to be a good parametric fit.

\begin{figure}
\plotone{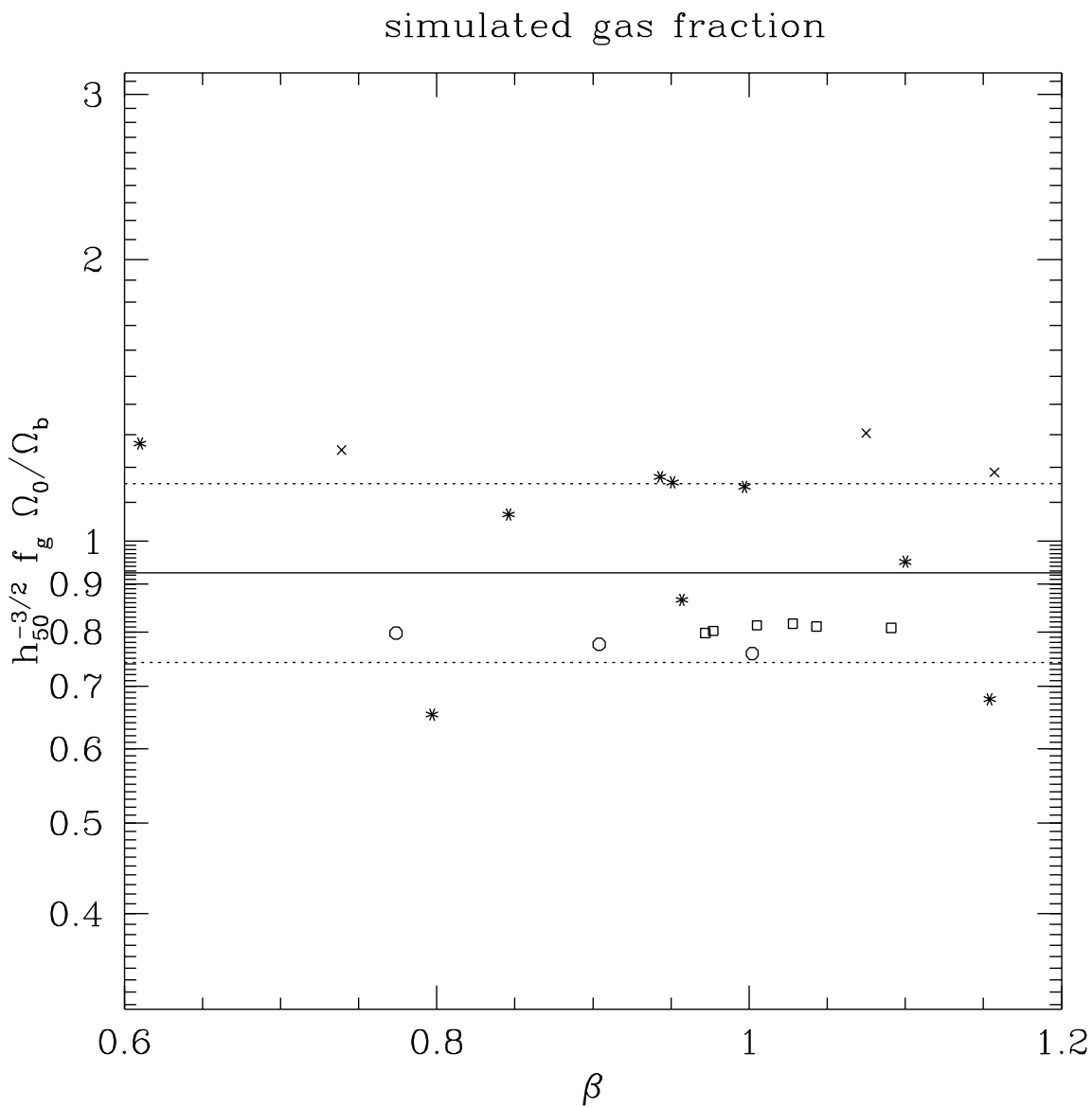}
\caption{The gas fractions inferred from simulations.  Each cluster is
projected, centered, and reduced in analogy to the real observed
clusters.  The solid line is the mean, and the dotted line the
standard deviation in the estimates.  No measurement errors are
included, implying that the scatter and mean reflect the intrinsic
scatter and bias of the fitting formula used in this paper.  Open
boxes are from model CDM.  Crosses are LCDM, asterixes are OCDM,
circles PREHEAT.}
\label{fig:simfg}
\end{figure}

We can repeat this exercise for each of the three possible
projections, and compile a database of inferred gas fractions.  The
result is shown in figure \ref{fig:simfg}.  The assumptions of
spherical symmetry, gas-tracing-mass, and hydrostatic equilibrium
actually work suprisingly well.  The reconstructed gas fraction has a
scatter of only 25\%, and the mean ratio is almost unbiased, differing
from the true mean by less than 8\%.

\section{Observations}

We base our local observational sample on the data set of JF98.  Their
catalog provides the X-ray luminosities, $\beta$, $r_c$ and $T$, just
like the simulated sample.  From these observables, we again construct
the cluster gas fractions.  The dominant error is generally the error
in temperature, which we took as the observational error on the gas
fraction.  In quadrature, we add the expected intrinsic 25\% error on
the gas fraction seen in the simulations in the absence of measurement
error.  This scatter arises from the imperfection in the simplifying
assumptions.  We use this total error to weight each observation.  In
addition, we impose certain cuts in the model.  We require $T_X>4$ keV
and $\beta>0.6$.  The first restriction is to eliminate potential
problems with non-gravitational heat sources, such as supernova
heating (Loewenstein and Mushotzky 1997).  Photoheating is always
negligible.  The cut in $\beta$ is necessary for two reasons.  In the
JF98, clusters where $\beta$ could not be directly measured were
assigned $\beta=0.6$, and we felt it important to discard clusters
where parameters were based on guesses instead of actual measurements.
An additional reason is to obtain a fair estimator of the total
cluster temperature (Pen 1997).  For small values of $\beta$, much of
the X-ray emission arises from the outer regions, and the temperature
errors are dominated by radii where significant temperature gradients
are expected.  Formally, the errors are infinite for $\beta\leq 4/7$.
Of the 57 clusters in the JF98 sample with measured temperatures, 25
pass these two critera.  We then apply equation (\ref{eqn:fg}) to the
clusters and take a weighted average.

The high redshift sample in the literature is still very meager.  A
search for clusters with redshift $z>0.5$ satisfying our requirements
turned up 3 clusters, MS0451, MS1054 and CL0016 (Donahue \etal 1997,
Mushotsky and Scharf 1997, Neumann and B\"ohringer 1996).  Their
properties and inferred gas fractions depend on their geometrical
distances.  For $q_0=1/2,\ \Omega=1,\ H_0=50$km/sec/Mpc, the results
are listed in table \ref{tbl:clust}.  To convert to other cosmological
models, we note that the constrained observables are flux $F$, angular
core size $\theta_c$, and temperature.  The inferred gas fraction in
equation (\ref{eqn:fg}) scales as $f_g
\propto F^{1/2} d_L \theta_c^{1/2} d_A^{1/2}/T_X^{5/4}$.  Recalling
that the angular diameter distance $d_A$ is related to the luminosity
distance $d_L$ by $d_L=(1+z)^2 d_A$, we can incorporate the only
cosmological dependence into $f_g
\propto d_A^{3/2}$.  We will consider the consequences in the next
section.

\begin{table}
\begin{center}
\begin{tabular}{|l|l|l|l|l|l|l|}
\hline
Cluster&$z$&$L_X$&        $T_X$     &$r_c$&$\beta$&$f_g$ \\
       &&($10^{44}$ erg/s)& (keV)    &kpc&       &      \\
\hline\hline
MS0451&0.55&41.7&$10.4^{+1.6}_{-1.3}$&280&1.01&$0.0843\pm 0.023$\\
MS1054&0.83&42.0&$14.7^{+4.6}_{-3.5}$&500&0.80&$0.0819\pm 0.020$\\
CL0016&0.54&30.3&$8.0^{+1}_{-1}     $&372&0.85&$0.136 \pm 0.036$\\
\hline
\end{tabular}
\end{center}
\caption{High redshift clusters used for this study.  The listed
quantities are reconstructed from observables by luminosity $L_X=4\pi
F d_L^2$ in terms of flux $F$ and luminosity diameter distance $d_L$,
while core radius $r_c=\theta_c d_A$ in terms of angular diameter
distance $d_A$ for $\Omega=1,\ H_0=50$.}
\label{tbl:clust}
\end{table}

\section{Results}

For the local cluster sample, the mean gas fraction is very well
determined.  Applying equation (\ref{eqn:fg}) to all qualifying
clusters, we find the raw gas fraction $f_g=0.14\pm 0.007
h_{50}^{-3/2}$.  If we correct for the likely 8\% bias in the
estimator as seen in the simulations, and add an equal error for
uncertainties in systematics due to Poisson limitations in the
simulations, we find
\beq
f_g=0.15\pm 0.01 h_{50}^{-3/2}
\eeq
at 1-$\sigma$ error bars, based on the 25 clusters JF98.  The result
is consistent with Evrard (1997) $f_g=0.17\pm 0.01 h_{50}^{-3/2}$
which uses an entirely different approach.  Our $\chi^2$ for the
assumed errors is $33.2$ for 24 degrees of freedom.  This is
consistent with the null hypothesis that all clusters have the same
intrinsic gas fraction, where all errors are solely due to measurement
errors and the 25\% scatter in the gas fraction estimator.  We have no
need to model any further sources of error.

\begin{figure}
\plotone{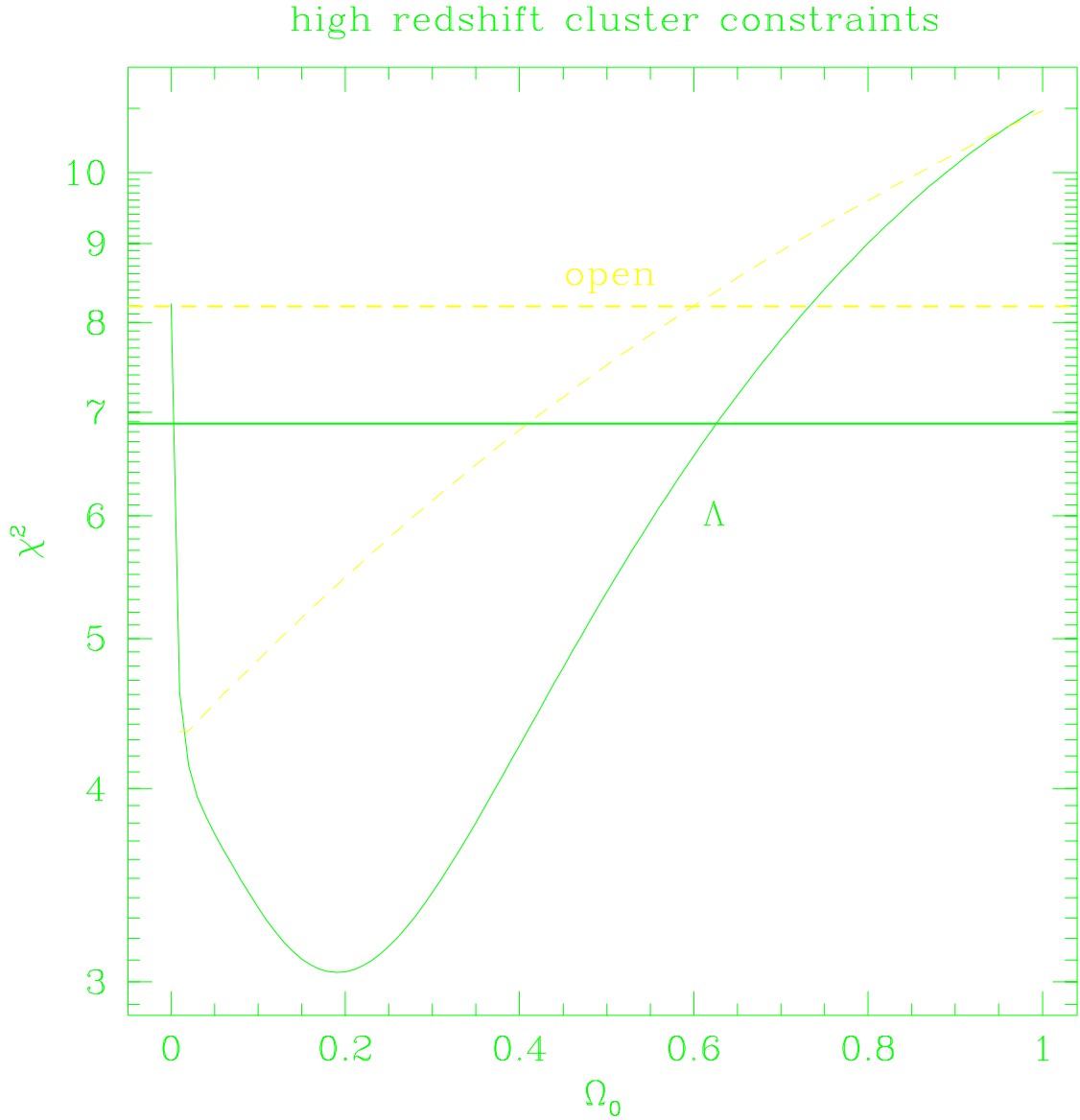}
\caption{$\chi^2$ as a function of the 
density parameter for hyperbolic (dashed) and flat $\Lambda$
cosmologies (solid).  The horizontal lines indicate the 5\%
likelihood cutoff for each model defined as $\Delta \chi^2=3.841$.}
\label{fig:chi2}
\end{figure}

The high redshift sample allows us to apply the fair sampling
hypothesis to constrain geometrical distances.  We define $\chi^2=\sum
(f_g^i - \bar{f_g})/\sigma_i^2$, where $\bar{f_g}=0.14$ and each gas
fraction and error is taken for the three high redshift clusters from
table \ref{tbl:clust}.  The individual gas fractions are a function of
cosmology $f_g^i=f_g^i(\Omega=1)\times
[d_A(\Omega_0,z)/d_A(\Omega_0=1,z)]^{3/2}$.  At a 95\% $\chi^2$ limit,
we obtain $-1<q_0<0.2$.  This interestingly does not include a flat
universe with $\Omega=1$.  We can quantify the $\chi^2$ distribution
for two specific cosmologies: a spatially flat universe with
cosmological constant $\Lambda$ and a spatially curved hyperbolic
universe.  In the first case, the angular diameter distance relation
is given by a simple fitting function (Pen 1998c):
\begin{eqnarray}
d_A &=&
\frac{c}{H_0(1+z)}\left[\eta(1,\Omega_0)-\eta(\frac{1}{1+z},\Omega_0)\right]
\nonumber\\
\eta(a,\Omega_0) &=& 2\sqrt{s^3+1}
\left[\frac{1}{a^4}-0.1540\frac{s}{a^3}+0.4304\frac{s^2}{a^2}
+0.19097\frac{s^3}{a}+0.066941s^4\right]^{-\frac{1}{8}}
\nonumber\\
s^3&=&\frac{1-\Omega_0}{\Omega_0}.
\label{eqn:dl}
\end{eqnarray}
In the open case, Mattig's relation gives
$d_A=2[2-\Omega_0+\Omega_0z-(2-\Omega_0)\sqrt{1-\Omega_0z}]/\Omega_0^2$
(Peebles 1993).  We then obtain the distribution of $\chi^2$ shown in
figure \ref{fig:chi2}.  For a spatially flat $\Lambda$ model, the
minimum $\chi^2=3.04$, which is good fit for 2 degrees of freedom.
We see the generic fact that high $\Omega_0$ universes are strongly
disfavored by these clusters.  In order to put statistically
meaningful limits, we need to understand the full distribution of
errors, which is clearly impractical.  But if we make the assumption
that errors are Gaussian, we obtain formal error estimates.  Varying
only $\Omega_0$, we obtain the 95\% confidence interval by requiring
$\Delta \chi^2=3.841$.  Formally, we find that $\Omega_0<0.63$ for a
spatially flat cosmological constant dominated universe, while
$\Omega_0<0.60$ for a spatially hyperbolic universe with no
cosmological constant, both at 95\% formal confidence.  This result is
consistent with the recent program of high-redshift supernova
observations (Perlmutter \etal 1998, Glanz 1998).  The hyperbolic
model is a questionable fit, since we limited the range of $\Omega_0$
to be positive for a meaningful geometry.

A short analysis of the difference to the Pen (1997) result is in
order.  That analysis obtained $q_0=0.85\pm 0.29$, which would argue
in favor of high $\Omega_0$.  CL0016 appears in both analyses, albeit
with slightly different gas fractions.  In the previous analysis, the
published central electron density $n_0$ was used, which is
unfortunately interpreted differently by various authors.  This
analysis uses solely robust observables, namely the flux, angular size,
and temperature, which are less subject to interpretational systematic
differences.

Several systematic errors can still change the conclusions.  The
cluster sample is not homogeneous, and may have selection effects.  The
intrinsic scatter in the observable is 25\%, which is comparable to the
expected effect at $z\sim 0.5$.  The signal comes partially from the
reduction in the error due to the large statistical sample, which
assumes independence and randomness of each error source.  We quote the
formal confidence intervals from the $\chi^2$ value by assuming
Gaussianity, which may not be a good assumption.

\section{Conclusions}

Comparing a catalog of local cluster properties with simulations, we
find that the data is consistent with a universally constant gas
fraction of $f_g=0.15\pm 0.01 h_{50}^{-3/2}$.  With the present day
uncertainty in $0.025 \lesssim \Omega_b h_{50}^2 \lesssim 0.1$
(Schramm and Turner 1997) and some errors in the Hubble constant, we
obtain no useful constaint on $\Omega_0$ using the low redshift
clusters.  Independent conservation of baryons and dark matter,
however, allows us to constrain $\Omega_0$ from the apparent evolution
of the cluster gas fraction.

We have shown that the 3 clusters with measured gas fractions at
$z>0.5$ are inconsistent with an $\Omega=1$ universe and the fair
sampling hypothesis.  For a spatially flat cosmological constant
dominated universe, we obtain a bound of $\Omega_0<0.63$ (95\%
formal confidence) with a best fit value of $\Omega_0=0.2$.  For a spatially
hyperbolic universe with only matter, we find $\Omega_0<0.60$ (95\%
formal confidence) with the maximum likelihood at $\Omega_0=0$.  The errors
are dominated by the temperature uncertainties in the high redshift
clusters, and future observations could reduce the errors by a factor
of two.

This work was supported by the National Science Foundation through REU
grant AST 9321943, NASA ATP grant TBD and the Harvard Milton Fund.
Computing time was provided by the National Center for Supercomputing
Applications.  We would like to thank Bill Forman and Christine Jones
for providing the cluster X-ray tables.

\begin{figure}
\plotone{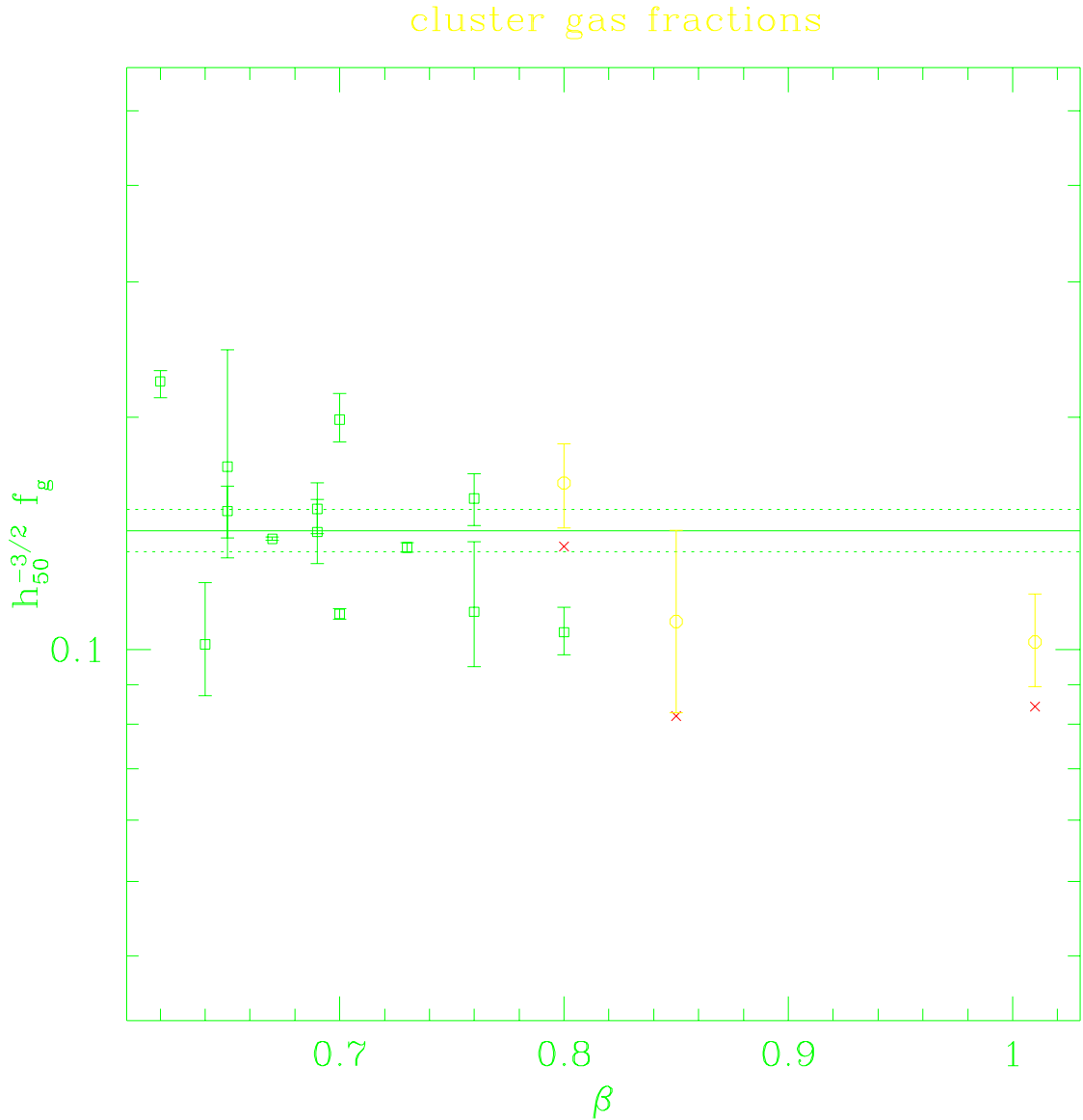}
\caption{The gas fractions of clusters with $T_X>4$ keV.  The error
bars are at 90\% confidence interval based on temperature errors
alone, and do not include a systematic 25\% scatter expected from
simulations.  The round points are the high redshift clusters for
$q_0=0$, while the crosses are the same clusters for $q_0=1/2$.  The
solid horizontal and dashed lines are the best fit mean and error in
the mean respectively.}
\label{fig:fghot}
\end{figure}

\end{document}